\documentstyle[12pt]{article}
\date{}
\author{Valerii Dryuma\thanks{Work supported in part by Grant RFFI, Russia-Moldova}\\[5mm]
{\it Institute of Mathematics and Informatics, AS RM,}\\[3mm] {\it
5 Academiei Street, 2028 Kishinev, Moldova},\\[3mm]{\it e-mail:
valery@dryuma.com; cainar@mail.md} }
\title{On the Equations of Nonstationary \\[1mm]Transonic Gas Flows}

\begin{document}
\maketitle
\date{}
\maketitle
\begin{abstract}
\ \ \ \ The examples of solutions of the Equations of
Nonstationary Transonic Gas Flows are considered.
 Their properties are discussed.
\end{abstract}


\medskip
\section{Introduction}

      Two-dimensional equation of Nonstationary Transonic Gas Flow has the form
\begin{equation}\label{dr:eq1}
2\,{\frac {\partial ^{2}}{\partial x\partial z}}f(x,y,z)+\left ({
\frac {\partial }{\partial x}}f(x,y,z)\right ){\frac {\partial
^{2}}{
\partial {x}^{2}}}f(x,y,z)-{\frac {\partial ^{2}}{\partial {y}^{2}}}f(
x,y,z) =0,
\end{equation}
 where variable $z$ is considered as the time-variable .

     Three-dimensional generalization of this equation is defined by the equation
     \begin{equation}\label{dr:eq2}
2\,{\frac {\partial ^{2}}{\partial x\partial z}}f(x,y,z,s)+\left
({ \frac {\partial }{\partial x}}f(x,y,z,s)\right ){\frac
{\partial ^{2}} {\partial {x}^{2}}}f(x,y,z,s)-{\frac {\partial
^{2}}{\partial {y}^{2}} }f(x,y,z,s)-\]\[-{\frac {\partial
^{2}}{\partial {s}^{2}}}f(x,y,z,s)  =0.
\end{equation}
 where variable $z$ here is considered as the time-variable.

     The solutions of these equations and a corresponding  bibliography have been considered recently in
     [1]

      In this article we apply the method of solution of the p.d.e.'s
     described first in [2] and developed then in [3],~[4].

      This method allow us to construct particular solutions of the partial nonlinear
       differential equation
\begin{equation}\label{Dr3}
F(x,y,z,f_x,f_y,f_z,f_{xx},f_{xy},f_{xz},f_{yy},f_{yz},f_{xxx},f_{xyy},f_{xxy},..)=0.
 \end{equation}
with the help of transformation of the function  and variables.

      Essence of method consists in a following presentation of the functions and variables
\begin{equation}\label{Dr4}
f(x,y,z,s)\rightarrow u(x,t,z,s),\quad y \rightarrow
v(x,t,z,s),\quad f_x\rightarrow u_x-\frac{u_t}{v_t}v_x,\quad
f_s\rightarrow u_s-\frac{u_t}{v_t}v_s,\]\[ f_z\rightarrow
u_z-\frac{u_t}{v_t}v_z,\quad f_y \rightarrow \frac{u_t}{v_t},
\quad f_{yy} \rightarrow \frac{(\frac{u_t}{v_t})_t}{v_t}, \quad
f_{xy} \rightarrow \frac{(u_x-\frac{u_t}{v_t}v_x)_t}{v_t},...
\end{equation}
where variable $t$ is considered as parameter.

  Remark that conditions of the type
   \[
   f_{xy}=f_{yx},\quad f_{xz}=f_{zx},\quad f_{xs}=f_{sx}...
   \]
are fulfilled at the such type of presentation.

  In result instead of equation (\ref{Dr3}) one get the
  relation between the new variables $u(x,t,z)$ and $v(x,t,z)$ and
  their partial derivatives
\begin{equation}\label{Dr5}
\Psi(u,v,u_x,u_z,u_t,u_s,v_x,v_z,v_t,v_s...)=0.
  \end{equation}

    This relation coincides with initial p.d.e at the condition $v(x,t,z,s)=t$
    and takes more general form after presentation of the functions $u,v$ in form $u(x,t,z,,s)=F(\omega,\omega_t...)$
    and $v(x,t,z,s)=\Phi(\omega,\omega_t...)$ with some function $\omega(x,t,z,s)$  .

    Example.

    The equation of Riemann wave
    \[
    {\frac {\partial }{\partial x}}f(x,y)+f(x,y){\frac {\partial }{
\partial y}}f(x,y)=0
\]
after $(u,v)$-transformation takes the form
\[
\left ({\frac {\partial }{\partial x}}u(x,t)\right ){\frac
{\partial } {\partial t}}v(x,t)-\left ({\frac {\partial }{\partial
t}}u(x,t) \right ){\frac {\partial }{\partial
x}}v(x,t)+u(x,t){\frac {\partial } {\partial t}}u(x,t)=0.
\]

    The substitution here of the expressions
    \[v(x,t)=t{\frac {\partial }{\partial t}}\omega(x,t)-\omega(x,t)
,\quad u(x,t)={\frac {\partial }{\partial t}}\omega(x,t)
\]
 give us the linear equation
\[{\frac {\partial }{\partial x}}\omega(x,t)+{\frac {\partial }{
\partial t}}\omega(x,t)=0
\]
with general solution
\[
\omega(x,t)={\it \_F1}(t-x),
\]
where ${\it \_F1}(t-x)$ is arbitrary function.

   Choice of the function ${\it \_F1}(t-x)$ and elimination of the parameter $t$ from the relations
   \[y-t\mbox {D}({\it \_F1})(t-x)+{\it \_F1}(t-x)=0,\quad f(x,y)-\mbox {D}({\it \_F1})(t-x)=0
\]
lead to the function $f(x,y)$ satisfying the Riemann wave
equation.

\medskip
\section{Two-dimensional case}

   The equation (\ref{dr:eq1}) after applying $(u,v)$- transformation
   with conditions
        \[
    u(x,t,z)=t{\frac {\partial }{\partial t}}\omega(x,t,z)-\omega(x,t,z)
\]
and \[ v(x,t,z)={\frac {\partial }{\partial t}}\omega(x,t,z)
\]
takes the form
\begin{equation}\label{Dr6}
\left ({\frac {\partial ^{2}}{\partial
{t}^{2}}}\omega(x,t,z)\right ) \left ({\frac {\partial }{\partial
x}}\omega(x,t,z)\right ){\frac {
\partial ^{2}}{\partial {x}^{2}}}\omega(x,t,z)-1-2\,\left ({\frac {
\partial ^{2}}{\partial {t}^{2}}}\omega(x,t,z)\right ){\frac {
\partial ^{2}}{\partial x\partial z}}\omega(x,t,z)+\]\[+2\,\left ({\frac {
\partial ^{2}}{\partial t\partial x}}\omega(x,t,z)\right ){\frac {
\partial ^{2}}{\partial t\partial z}}\omega(x,t,z)-\left ({\frac {
\partial ^{2}}{\partial t\partial x}}\omega(x,t,z)\right )^{2}{\frac {
\partial }{\partial x}}\omega(x,t,z)
=0.
\end{equation}

  Its solution of the form
\[
\omega(x,t,z)=A(t,z)-xB(t)
\]
lead to the equation on the function $A(t,z)$
\[-1-2\,\left ({\frac {d}{dt}}B(t)\right
){\frac {\partial ^{2}}{
\partial t\partial z}}A(t,z)+\left ({\frac {d}{dt}}B(t)\right )^{2}B(t
)=0
\]
with solution
\[A(t,z)={\it \_F2}(t)+{\it \_F1}(z)+\left (-1/2\,\int \!\left ({\frac {d}{dt}}
B(t)\right )^{-1}{dt}+1/4\,\left (B(t)\right )^{2}\right )z
\]
 where ${\it \_F2}(t),~B(t),~{\it \_F1}(z)$ are arbitrary functions.

    In result we find that the function
    \[
    \omega(x,t,z)={\it \_F2}(t)+{\it \_F1}(z)+\left (-1/2\,\int \!\left ({
\frac {d}{dt}}B(t)\right )^{-1}{dt}+1/4\,\left (B(t)\right )^{2}
\right )z-xB(t)
\]
is the solution of the equation (\ref{Dr6}).

     After the choice of the functions ${\it \_F2}(t),~B(t)$ and  elimination of the parameter $t$ from the relations
        \[
    f-\left(t{\frac {\partial }{\partial
    t}}\omega(x,t,z)-\omega(x,t,z)\right)=0
\quad y-{\frac {\partial }{\partial t}}\omega(x,t,z) =0
\]
one gets the solution of the equation (\ref{dr:eq1}).

   Let us consider some examples.

   In the case
   \[
   {\it \_F2}(t)=0,\quad B(t)={t}^{-1},
\]
we find the relations
\[
4\,f(x,y,z){t}^{2}-4/3\,z{t}^{5}-8\,xt+3\,z+4\,{\it
\_F1}(z){t}^{2}=0,
\]
and
\[
2\,y{t}^{3}-z{t}^{5}-2\,xt+z=0.
\]

   Elimination of the parameter $t$ from these relations in the case
${\it \_F1}(z)=0$ lead to the solution  $f(x,y,z)$ of the equation
(\ref{dr:eq1}) satisfying the algebraic equation
\[
248832\,\left (f(x,y,z)\right )^{5}{z}^{2}-248832\,{x}^{2}\left
(f(x,y ,z)\right )^{4}z+\left
(-1451520\,{z}^{2}xy-221184\,{y}^{3}z\right ) \left
(f(x,y,z)\right )^{3}+\]\[+\left
(-216000\,{z}^{4}x+475200\,{z}^{3}{y
}^{2}+1327104\,yz{x}^{3}+221184\,{y}^{3}{x}^{2}\right )\left
(f(x,y,z) \right )^{2}+\]\[+\left
(90000\,y{z}^{5}+995328\,{y}^{4}zx+614400\,{z}^{3}{
x}^{3}+1290240\,{y}^{2}{z}^{2}{x}^{2}\right
)f(x,y,z)-\]\[-518400\,{y}^{3}{
z}^{3}x-393216\,{x}^{5}{z}^{2}-1179648\,{x}^{4}{y}^{2}z+3125\,{z}^{7}-
373248\,{y}^{5}{z}^{2}-\]\[-192000\,{z}^{4}{x}^{2}y-884736\,{y}^{4}{x}^{3}=0.
\]

\medskip
\section{Three-dimensional generalization}

   In a three dimensional case the equation  of Nonstationary Transonic Gas Flow
   takes the form
\begin{equation}\label{Dr7}
2\,{\frac {\partial ^{2}}{\partial x\partial z}}f(x,y,z,s)+\left
({ \frac {\partial }{\partial x}}f(x,y,z,s)\right ){\frac
{\partial ^{2}} {\partial {x}^{2}}}f(x,y,z,s)-\]\[-{\frac
{\partial ^{2}}{\partial {y}^{2}} }f(x,y,z,s)-{\frac {\partial
^{2}}{\partial {s}^{2}}}f(x,y,z,s) =0.
\end{equation}

    Recall that the variable $z$ in this equation play the role
    of a time-variable.

     After application of $u,v$-transformation of the form
\[
u(x,t,z,s)=t{\frac {\partial }{\partial
t}}\omega(x,t,z,s)-\omega(x,t, z,s) ,\quad v(x,t,z,s)={\frac
{\partial }{\partial t}}\omega(x,t,z,s)
\]
we find from (\ref{Dr7}) the equation
\begin{equation}\label{Dr8}
\left ({\frac {\partial ^{2}}{\partial
{t}^{2}}}\omega(x,t,z,s)\right ){\frac {\partial ^{2}}{\partial
{s}^{2}}}\omega(x,t,z,s)-2\,\left ({ \frac {\partial
^{2}}{\partial {t}^{2}}}\omega(x,t,z,s)\right ){\frac {\partial
^{2}}{\partial x\partial z}}\omega(x,t,z,s)+\]\[+\left ({\frac {
\partial ^{2}}{\partial {t}^{2}}}\omega(x,t,z,s)\right )\left ({\frac
{\partial }{\partial x}}\omega(x,t,z,s)\right ){\frac {\partial
^{2}}{
\partial {x}^{2}}}\omega(x,t,z,s)-\]\[-\left ({\frac {\partial ^{2}}{
\partial t\partial x}}\omega(x,t,z,s)\right )^{2}{\frac {\partial }{
\partial x}}\omega(x,t,z,s)+2\,\left ({\frac {\partial ^{2}}{\partial
t\partial x}}\omega(x,t,z,s)\right ){\frac {\partial
^{2}}{\partial t
\partial z}}\omega(x,t,z,s)-\]\[-1-\left ({\frac {\partial ^{2}}{\partial s
\partial t}}\omega(x,t,z,s)\right )^{2}=0.
\end{equation}

   From here in the case
    \[\omega(x,t,z,s)=A(t,s)+k\left (x+z\right )t
\]
one gets the Monge-Ampere  equation on the function $A(t,s)$
\begin{equation}\label{dr:eq2}
1-\left ({\frac {\partial ^{2}}{\partial {t}^{2}}}A(t,s)\right ){
\frac {\partial ^{2}}{\partial {s}^{2}}}A(t,s)+\left ({\frac {
\partial ^{2}}{\partial s\partial t}}A(t,s)\right )^{2}
 =0.
\end{equation}

    It is possible to show that the equation
(\ref{dr:eq2}) can be
   integrated with the help of corresponding $(u,v)$-transformation.

   Its solutions are dependent from solutions of the
   linear Laplace equation.

   In fact, the equation
   \[
   \left ({\frac {\partial ^{2}}{\partial {x}^{2}}}f(x,y)\right ){\frac {
\partial ^{2}}{\partial {y}^{2}}}f(x,y)-\left ({\frac {\partial ^{2}}{
\partial x\partial y}}f(x,y)\right )^{2}-1=0
\]
after $(u,v)$-transformation with
\[
u(x,t)=t\omega_t-\omega,\quad v(x,t)=\omega_t
\]
 takes the form of linear Laplace equation
 \[{\frac {\partial ^{2}}{\partial {x}^{2}}}\omega(x,t)+{\frac {\partial
^{2}}{\partial {t}^{2}}}\omega(x,t)=0
\]
and its particular solutions after elimination of parameter $t$
give us the solutions of the Monge-Ampere equation (\ref{dr:eq2}).

     The substitution of another form
     \[\omega(x,t,z,s)=A(t,z)+\left (s+x\right )t
\]
into the equation (\ref{Dr8}) lead to the equation on the function
$A(t,z)$
\[
-2-t+2\,{\frac {\partial ^{2}}{\partial t\partial z}}A(t,z)=0
\]
having the general solution
\[
A(t,z)={\it \_F2}(t)+{\it \_F1}(z)+tz+1/4\,{t}^{2}z
\]
where ${\it \_F2}(t),~{\it \_F1}(z)$ are arbitrary functions.

    The choice of the functions ${\it \_F2}(t),~{\it \_F1}(z)$
    allow us to construct solutions of initial equation.

    For example in the case
    \[{\it \_F2}(t)={t}^{-1}
,\quad {\it \_F1}(z)=0
\]
 elimination of the parameter $t$ from the relations
\[
4\,f(x,y,z,s)t-{t}^{3}z+8=0 ,\quad
2\,y{t}^{2}-2\,{t}^{2}z-{t}^{3}z-2\,{t}^{2}s-2\,{t}^{2}x+2 =0
\]
give us the solution of the equation (\ref{Dr7}) satisfying the
algebraic equation
\[
16\,\left (f(x,y,z,s)\right )^{3}z+\]\[+\left
(32\,zy-32\,zx-32\,zs-32\,sx+
32\,sy-16\,{y}^{2}-16\,{z}^{2}-16\,{s}^{2}-16\,{x}^{2}+32\,yx\right
) \left (f(x,y,z,s)\right )^{2}+\]\[+\left
(72\,zy-72\,zx-72\,{z}^{2}-72\,zs \right
)f(x,y,z,s)+192\,z{s}^{2}+64\,{s}^{3}-64\,{y}^{3}-\]\[-384\,zyx+384
\,zsx-384\,zys+27\,{z}^{2}+192\,s{x}^{2}+64\,{x}^{3}+64\,{z}^{3}+192\,
{z}^{2}s+192\,{z}^{2}x+192\,z{y}^{2}+\]\[+192\,s{y}^{2}+192\,{y}^{2}x-192\,
y{z}^{2}-192\,y{s}^{2}-192\,y{x}^{2}+192\,z{x}^{2}+192\,{s}^{2}x-384\,
ysx=0.
\]

\section{ The case of axisymmetric equation}

   The equation (\ref{Dr7})
         \[
   2\,{\frac {\partial ^{2}}{\partial x\partial y}}f(x,y,z,s)+\left ({
\frac {\partial }{\partial x}}f(x,y,z,s)\right ){\frac {\partial
^{2}}{
\partial {x}^{2}}}f(x,y,z,s)-\Delta f(x,y,z,s)=0,
\]
where
\[\Delta=\frac{\partial}{\partial z^2}+\frac{\partial}{\partial
s^2},
\]
   in polar coordinates $s=r\cos(\phi),~z=r\sin(\phi)$~  takes the form
\begin{equation}\label{Dr9}
   2\,{\frac {\partial ^{2}}{\partial x\partial y}}f(x,y,z)+\left ({
\frac {\partial }{\partial x}}f(x,y,z)\right ){\frac {\partial
^{2}}{
\partial {x}^{2}}}f(x,y,z)-{\frac {\partial ^{2}}{\partial {z}^{2}}}f(
x,y,z)-{\frac {{\frac {\partial }{\partial z}}f(x,y,z)}{z}} =0
\end{equation}
 where we replace variable $r$ on variable $z$

  After applying $(u,v)$-transformation with

    \[
    v(x,t,z)=t{\frac {\partial }{\partial t}}\omega(x,t,z)-\omega(x,t,z)
,\quad u(x,t,z)={\frac {\partial }{\partial t}}\omega(x,t,z)
\]
the equation (\ref{Dr9}) takes the form
\[
-\left ({\frac {\partial ^{2}}{\partial
{t}^{2}}}\omega(x,t,z)\right ) z{t}^{2}\left ({\frac {\partial
}{\partial x}}\omega(x,t,z)\right ){ \frac {\partial
^{2}}{\partial {x}^{2}}}\omega(x,t,z)+{t}^{3}\left ({ \frac
{\partial }{\partial z}}\omega(x,t,z)\right ){\frac {\partial ^{
2}}{\partial {t}^{2}}}\omega(x,t,z)+\]\[+\left ({\frac {\partial
^{2}}{
\partial {t}^{2}}}\omega(x,t,z)\right ){t}^{3}z{\frac {\partial ^{2}}{
\partial {z}^{2}}}\omega(x,t,z)-{t}^{3}z\left ({\frac {\partial ^{2}}{
\partial t\partial z}}\omega(x,t,z)\right ){\frac {\partial ^{2}}{
\partial {z}^{2}}}\omega(x,t,z)+\]\[+{t}^{2}z\left ({\frac {\partial }{
\partial z}}\omega(x,t,z)\right ){\frac {\partial ^{2}}{\partial {z}^{
2}}}\omega(x,t,z)+z{t}^{2}\left ({\frac {\partial ^{2}}{\partial t
\partial x}}\omega(x,t,z)\right )^{2}{\frac {\partial }{\partial x}}
\omega(x,t,z)-\]\[-2\,zt\left ({\frac {\partial ^{2}}{\partial
t\partial x} }\omega(x,t,z)\right )\left ({\frac {\partial
}{\partial x}}\omega(x,t ,z)\right )^{2}+z\left ({\frac {\partial
}{\partial x}}\omega(x,t,z) \right )^{3}+2\,zt{\frac {\partial
}{\partial x}}\omega(x,t,z)-\]\[-2\,z{t} ^{2}{\frac {\partial
^{2}}{\partial t\partial x}}\omega(x,t,z) =0.
\]

   This equation has solution of the form
   \[\omega(x,t,z)=A(x,z)t+B(t)
\]
where function  $B(t)$ is arbitrary and the function $A(x,z)$
satisfies the equation
\begin{equation}\label{Dr11}
z\left ({\frac {\partial }{\partial x}}A(x,z)\right ){\frac
{\partial ^{2}}{\partial {x}^{2}}}A(x,z)-{\frac {\partial
}{\partial z}}A(x,z)-z {\frac {\partial ^{2}}{\partial
{z}^{2}}}A(x,z) =0. \end{equation}

    The solutions of the equation (\ref{Dr11}) can be obtained with
    the help of $(u,v)$-transformation and a simplest of them
    has the form
    \[
    A(x,z)={\it \_C2}-{\it \_C1}-{\it \_C2}\,\ln ({\frac {{\it \_C2}}{z}})
-x.
\]

   Using the expression
   \[\omega(x,t,z)=A(x,z)t+B(t)
\]
with a given function $A(x,z)$ and arbitrary function $B(t)$
the solution of the equation (\ref{Dr9}) can be constructed.

   As example, in the case $B(t)=\ln(t)$ the elimination of the
   parameter $t$ from the relations
   \[f(x,y,z)t-t{\it \_C2}+t{\it \_C1}+t{\it \_C2}\,\ln ({\frac {{\it \_C2}
}{z}})+tx-1=0,\]\[y-1+\ln (t)=0
\]
lead to the solution of the equation
(\ref{Dr9})
\[f(x,y,z)=-\left(-{e^{-y+1}}{\it \_C2}+{e^{-y+1}}{\it
\_C1}+{e^{-y+1}}{\it \_C2} \,\ln ({\frac {{\it
\_C2}}{z}})+{e^{-y+1}}x-1\right)\left ({e^{-y+1}} \right )^{-1}.
\]

    In the case $B(t)=t{e^{t}}$ by analogy way we find the
    solution
    \[
    f(x,y,z)=-{\it \_C1}-{\it \_C2}\,\ln ({\frac {{\it \_C2}}{z}})-x+{\it \_C2}+1/2
\,{\frac {y}{{\it LambertW}(1/2\,\sqrt {y})}}+\]\[+1/4\,{\frac
{y}{\left ({ \it LambertW}(1/2\,\sqrt {y})\right )^{2}}}.
\]

    The solution of the equation (\ref{Dr11})
\[
A(x,z)=2/9\,{\frac {{x}^{3}}{{z}^{2}}}
\]
lead to  the function
\[
\omega(x,t,z)=2/9\,{\frac {{x}^{3}t}{{z}^{2}}}+B(t)
\]
where $B(t)$ is arbitrary function.

    In the case $B(t)=\ln (t)$ we find
    \[
    f(x,y,z)=2/9\,{\frac {{x}^{3}}{{z}^{2}}}+\left ({e^{-y+1}}\right
    )^{-1}.
\]

   In the case $B(t)={t}^{2}+{t}^{-2}$ we get
   \[
   f(x,y,z)=2/9\,{\frac {\left (\sqrt {2\,y+2\,\sqrt {{y}^{2}+12}}\sqrt {{y}^{2}+
12}+\sqrt {2\,y+2\,\sqrt {{y}^{2}+12}}y\right ){x}^{3}}{\left (y+
\sqrt {{y}^{2}+12}\right )\sqrt {2\,y+2\,\sqrt
{{y}^{2}+12}}{z}^{2}}}+ \]\[+2/9\,{\frac
{72\,{z}^{2}+18\,{z}^{2}{y}^{2}+18\,{z}^{2}y\sqrt {{y}^{2}
+12}}{\left (y+\sqrt {{y}^{2}+12}\right )\sqrt {2\,y+2\,\sqrt
{{y}^{2} +12}}{z}^{2}}}.
\]

\medskip
\section{Acknowledgement}

   This research was partially supported  by the GRant 06.01 CRF of HCSTD ASM and
    the RFBR Grant

 {\small
\centerline{\bf References:}

\smallskip
\noindent 1. Xiaoping Xu,  {\it Stable-Range Approach to the
Equation of Nonstationary Transonic Gas flows},~ ArXiv: 0706.4189
v1 , physics.flu-dyn, 28 Jun 2007, p. 1-19.

\smallskip
\noindent 2. V. Dryuma, {\it The Riemann and Einsten-Weyl
geometries in theory of differential equations, their applications
and all that}. A.B.Shabat et all.(eds.), New Trends in
Integrability and Partial Solvability, Kluwer Academic Publishers,
Printed in the Netherlands , 2004, p.115--156.}

\smallskip
\noindent 3. Dryuma V.S., {\it On solutions of the heavenly
equations and their generalizations},  ArXiv:gr-qc/0611001 v1 31
Oct 2006, p.1-14.

\smallskip
\noindent 3. Dryuma V.S., {\it On dual equation in theory of the
second order ODE's},  ArXiv:nlin/0001047 v1 22 Jan 2007, p.1-17.

\end{document}